\documentclass[12pt]{article}
\usepackage{latexsym}
\begin{document}
\def\ba{\begin{array}}
\def\ea{\end{array}}
\def\be{\begin{equation}}
\def\ee{\end{equation}}
\def\bem{\begin{em}}
\def\eem{\end{em}}
\def\ot{\otimes}
\def\ld{\ldots}
\def\c{\times}
\def\a{\alpha}
\def\b{\beta}
\def\d{\delta}
\def\D{\Delta}
\def\e{\varepsilon}
\def\vp{\varphi}
\def\ra{\longrightarrow}
\def\M{\cal M}
\def\ca{{\cal A}}
\def\cb{{\cal B}}
\def\cc{{\cal C}}
\def\cd{{\cal D}}
\def\cm{{\cal M}}
\def\cv{{\cal V}}
\def\cw{{\cal W}}
\def\dn#1,{_{({#1})}}
\def\real{I \kern-.33em R}
\def\com{I \kern-.66em  C}
\renewcommand{\thesection}{\Roman{section}}
\def\pr{(.|.)}
\def\lb{\langle}
\def\rb{\rangle}
\def\R{\hat{R}}
\def\m#1,{m_{#1}}
\def\ic#1,{i_{#1}}
\def\jc#1,{j_{#1}}
\def\tt#1,{>\!\!\!\lhd_{#1}}
\def\tw{>\!\!\!\lhd}
\def\hs{\hspace{0.3cm}}
\def\op{\mathrm{op}}
\title{ON CROSSED PRODUCT OF ALGEBRAS\thanks{
to be published in J. Math. Phys. v41 no10 (2000)} }
\author{Andrzej Borowiec and W{\l}adys{\l}aw Marcinek
\\Institute of Theoretical Physics, University of Wroc{\l}aw
\\Plac Maxa Borna 9,  50-204 Wroc{\l}aw
\\POLAND}
\date{}
\maketitle
\begin{abstract}
The concept of a crossed tensor product of algebras is studied
from a few points of views. Some related constructions are
considered. Crossed enveloping algebras and their representations
are discussed. Applications to the noncommutative geometry and
particle systems with generalized statistics are indicated.
\end{abstract}

\vspace{3cm}

PACS. 02. 40. +m - Differential geometry in theoretical physics.\\
PACS. 03. 65. Fd - Algebraic methods in quantum theory
\newpage
\section{Introduction}
The notion of a crossed product of Hopf algebras is well--known
\cite{mon}. It is also known that there is an algebra analogue for
such product called a crossed (braided) product of algebras. It
has been used in several constructions in the area of the
nonocommutative geometry, quantum groups and braided categories.
For example this product in braided categories has been studied
intensively by Majid \cite{sm,sm2,sm3,sm4}. It is interesting that
there is a more general notion of a crossed product of algebras
without the notion of the braided categories. Namely, if $\ca$ and
$\cb$ are two unital and associative algebras over a field $k$,
then such product is formed by a bigger algebra $\cw$. This
algebra contains algebras $\ca$ and $\cb$ as subalgebras in such a
way that $\cw$ is generated as an algebra just by $\ca$ and $\cb$.
This product is called in general a crossed (or a twisted) tensor
product and it has been recently studied on an abstract algebraic
level by Van Daele and Van Keer \cite{dk}, by {\v C}ap, Schichl
and Van{\v zura} \cite{csv}. An application of such product in the
area of $C^{\ast}$--algebras has been considered by Woronowicz
\cite{wor}. A crossed tensor product has been also used by
Zakrzewski \cite{zak} in the study of quantum Lorentz and
Poincar\'e groups. An interesting approach for the study of
noncommutative de Rham complexes has been developed by Manin
\cite{man2}. It is based on the notion of the so--called skew
product of algebras. Similar concept corresponding to the algebra
of differential forms on a full matrix bialgebra has been
developed by Sudbery \cite{sud}. According to his construction
such algebra is a skew product of an algebra of functions and an
algebra of differential forms with constant coefficients. It is
obvious that such skew product provide an example of a crossed
tensor product. Related subject has been also considered by Wambst
\cite{wam}. One can see that in general the algebra $\cw$ of
differential operators acting on an algebra $\ca$ can be described
as a crossed product of the algebra and the algebra of vector
fields corresponding for an arbitrary noncommutative differential
calculus \cite{bor,BKO2,BK,BKO3}.

In the present paper we are going to study the concept of a
crossed tensor product of algebras from a few different points of
views: module theory, Hopf algebras, free product of algebras and
some constructions related to the noncommutative geometry. Our
considerations are motivated by the application to the
construction of the so--called crossed enveloping algebras and
their representations. Note that Wick algebras considered
previously in the study of deformed commutation relations
\cite{js} are particular examples of crossed enveloping algebras.
Such algebras has been also important in the study of systems with
generalized quantum statistics \cite{mar3,mar4,qstat}.

The paper is organized as follows. We recall the definition of the
crossed product in the Section 2. The corresponding module structures
are considered in this section. The relation between this product
and the smash product or the semi--simple product of Hopf algebras is
given. The connection with the free product of algebras is studied in
Section 3. The construction of twisted product for free algebras is
described in details in the Section 4. Ideals in the twisted products
and corresponding quotient constructions are studied in the Section 5.
Consistency conditions for such constructions are described as
consequences of axioms for the twisted product. Some examples are
given. In the Section 6 representations of the twisted tensor
product are considered. Crossed enveloping algebras are described as
a twisted tensor product of a pair of conjugated algebras.
Representations of crossed enveloping algebras are also considered.
As an example the Fock space representation for a system with
generalized statistics is given.
\section{Preliminaries}
In this note $k$ is a field of complex (or real) numbers . All
objects considered here  are first of all $k$-vector spaces. All
maps are assumed to be $k$-linear maps. The tensor product $\ot$
means $\ot_{k}$. In what follows algebra means associative unital
$k$-algebra and homomorphisms are assumed to be unital. If $\ca$
is an algebra then $\ca^\op$ denotes algebra with the opposite
multiplication: $a\cdot_\op a^\prime=a^\prime\,a$ . For
comultiplication $\Delta$ we shall use a shorthand Sweedler
(sigma) notation $\Delta (a)= \sum_i a^{'}_{i}\ot a^{''}_{i}\equiv
a^{(1)}\ot a^{(2)}$  (with $\sum_i$ omitted). Likewise, throughout
the paper we will use the Sweedler type notation for a twisting
map (see below)\ $\tau :\cb\ot\ca\ra\ca\ot\cb$ :i.e. we will write
$\tau(b\ot a)\equiv a_{(1)}\ot b_{(2)}$;  again the summation is
assumed here but not written explicitly.

Let $\ca$ and $\cb$ be two unital and associative algebras over
$k$. The multiplication in these algebras is denoted by $\m \ca,$
and $\m \cb,$, respectively. Let us recall briefly the concept of
a crossed product of algebras \cite{csv}.

\bem
{\bf Definition:}
An associative algebra $\cw$ equipped with two injective algebra
homomorphisms $\ic \ca, :\ca\hookrightarrow\cw$ and
$\ic \cb, :\cb\hookrightarrow\cw$ such that the canonical linear
map
$\Phi : \ca \ot_k \cb \ra \cw$ defined by
\be
\Phi (a\ot b) := m_{\cw}\circ (\ic\ca,\ot\ic\cb,)(a\ot b)
\ee
is a linear isomorphism is said to be a crossed (twisted)  product of
$\ca$ and $\cb$.
\eem

The above definition means that the crossed product of algebras
$\ca$ and $\cb$ is a bigger algebra $\cw$ which contain these two
algebras as subalgebras in such a way that $\cw$ is generated by
$\ca$ and $\cb$. In particular as a linear space the algebra $\cw$
is isomorphic to $\ca\ot\cb$. The definition is given up to the
isomorphism of algebras. As an example, we can consider the
standard tensor product of algebras with multiplication given by
the formula $(a\ot b)(a^\prime\ot b^\prime):= aa^\prime\ot
bb^\prime$ for $a\ot b,\,a^\prime\ot b^\prime\in \ca\ot\cb$. For
our purposes here, we shall denote by $\ca\ot_c\cb$ an algebra
being the standard tensor product of two algebras $\ca$ and $\cb$.
One will see in the moment that this example does not exhaust all
possible cases.

First, let us study module structures on the above crossed product
of algebras. If $\ca$ and $\cb$ are algebras, then an arbitrary
linear space $\cm$ is said to be a $(\ca, \cb)$--bimodule if $\cm$
is left $\ca$--module and right $\cb$--module and both structures
commute, i. e.
\be
(a.m).b = a.(m.b),\quad a\in\ca, b\in\cb, m\in\cm . \ee In other
words, $(\ca, \cb)$--bimodules are left modules over
$\ca\ot_c\cb^\op$. We shall also identify $\ca$--bimodules with
$(\ca, \ca)$--bimodules. Observe that the tensor product
$\ca\ot\cb$ has a canonical left $\ca$--module and right
$\cb$--module structure defined by
\be
\ba{l} a.(a'\ot b) := aa'\ot b ,\quad (a\ot b).b' := a\ot bb' ,
\label{lms} \ea \ee respectively. This means that $\ca\ot\cb$
inherits a $(\ca, \cb)$--bimodule structure in natural way. The
$(\cb, \ca)$--bimodule structure on $\ca\ot\cb$ is a problem. We
use the concept of module cross \cite{BKO3} for the study of this
problem. Let $\tau : \cb\ot\ca\ra\ca\ot\cb$ be a linear mapping,
then we define the left, right and two-sided universal lift of
$\tau$ as mappings (cf. \cite{BKO3})
\be
\ba{ll} ^u\tau : \ca\ot\cb\ot\ca\ra\ca\ot\cb,& ^u\tau (a\ot b \ot
a') := a'.\tau(b\ot a),\\ \tau^u : \cb\ot\ca\ot\cb\ra\ca\ot\cb,&
\tau^u (b'\ot a \ot b) := \tau(b\ot a).b',\\ ^u\tau^u :
\cb\ot\ca\ot\cb\ot\ca\ra\ca\ot\cb,& ^u\tau^u (b'\ot a \ot b\ot a')
:= a'.\tau(b\ot a).b', \ea \ee respectively.

\bem
{\bf Lemma:}
A mapping $\tau : \cb\ot\ca\ra\ca\ot\cb$ define on $\ca\ot\cb$
a structure of:
(i) a right $\ca$--module, (ii) left $\cb$--module,
(iii) $(\ca, \cb)$--bimodule,
if and only if the corresponding universal lift $^u\tau$, $\tau^u$ or
$^u\tau^u$ is the algebra homomorphism
\be
\ba{ll} (i)&^u\tau\in alg(\ca^\op,\ End_{k}(\ca\ot\cb)),\\
(ii)&\tau^u \in alg(\cb,\ End_{k}(\ca\ot\cb)),\\ (iii)&^u\tau^u\in
alg(\cb\ot_c\ca^\op,\ End_{k}(\ca\ot\cb)), \ea \ee \eem
respectively.

{\bf Proof:} Let $\cm$ be a left $\cb$-module. For each $b\in\cb$
define $L_b\in End_k\cm$ by $L_b(m):= b.m$. Then $L:\cb\ra
End_k\cm$ is an algebra homomorphism. Similarly,  right
$\ca$-module structures on $\cm$ are in one-to-one correspondence
with left $\ca^\op$ structures on $\cm$. Now (iii) is obvious,
since any $(\cb, \ca)$-bimodule structure is in fact, due to
commutativity (2), a left $\cb\ot_c\ca^\op$ structure. \hfill
$\Box$\\

\bem
{\bf Definition:} A $(\ca, \cb)$--bimodule $\cw$ which is at the same
time a $(\cb, \ca)$--bimodule, $\ca$-bimodule and $\cb$--bimodule is
said to be a crossed $(\ca, \cb)$--bimodule.
\eem

\bem {\bf Theorem:} There is one to one correspondence between
crossed $(\ca, \cb)$--bimodule structure on $\ca\ot\cb$ and linear
mappings $\tau:\cb\ot\ca\ra\ca\ot\cb$ satisfying the following
relations
\be
\ba{l}
\tau (1 \ot a) = a \ot 1\\
\tau \circ (m_{\cb} \ot id_{\ca}) = (id_{\ca} \ot m_{\cb})
\circ (\tau \ot id_{\cb}) \circ (id_{\cb} \ot \tau)
\label{mtr}
\ea
\ee
and
\be
\ba{l}
\tau (b\ot 1) = 1 \ot b \\
\tau \circ (id_{\cb} \ot m_{\ca}) = (m_{\ca} \ot id_{\cb})
\circ (id_{\ca} \ot \tau) \circ (\tau \ot id_{\ca}),
\label{mtw}
\ea
\ee
\eem

{\bf Proof:}
The left $\ca$--module and a right $\cb$--module acting on $\ca\ot\cb$
is given by formulae (\ref{lms}). We define a right $\ca$--module and
a left $\cb$--module action on $\ca\ot\cb$ by formulae
\be
\ba{l} (a\ot b).a' := a'\tau (b\ot a) = \!^u\tau(a\ot b\ot a'),\\
b'.(a\ot b) := \tau (b\ot a)b' = \tau^u (b'\ot a\ot b), \ea \ee
respectively. \hfill $\Box$\\

\bem {\bf Definition:} A $k$-linear mapping $\tau :
\cb\ot\ca\ra\ca\ot\cb$ satisfying the condition (\ref{mtr}) is
called a left $\cb$-module cross. Similarly, if the relation
(\ref{mtw}) is satisfied, then $\tau$ is called a right
$\ca$-module cross. The map $\tau$ is said to be an {\it algebra
cross} if it is both a left $\cb$- and right $\ca$-module cross.
\eem

 It is obvious that the standard twist (switch) $\tau:\cb\ot\ca\ra\ca\ot\cb$
defined by $\tau (b\ot a) := a\ot b$ satisfies all conditions for
the cross. It give rise to the standard tensor product of algebras
$\ca\ot_c\cb$. The second example is a graded algebra version of
the previous one. It is provided by the following graded twist
\be
\tau (b\ot a) := (-1)^{mn} a\ot b , \ee where $a\in\ca,\ b\in\cb$
are homogeneous elements of graded algebras $\ca$ and $\cb$ of
grade $m$ and $n$, respectively. Let $\tau :
\cb\ot\ca\ra\ca\ot\cb$ be an algebra cross, then according to the
last theorem there exists a structure of a crossed $(\ca,
\cb)$--bimodule on $\ca\ot\cb$. This structure will be denoted by
$\ca\tw_{\tau}\cb$.

\bem
{\bf Lemma:}
Let $\cw$ be a crossed $(\ca , \cb)$--bimodule. Assume that the algebra
$\ca$ as $k$-submodule universally generates $\cw$ as
a left $\ca$--module and similarly $\cb$ generates $\cw$ as
a right $\cb$--module. Then there exist the unique algebra cross
$\tau :\cb\ot\ca\ra\ca\ot\cb$ such that $\cw = \ca\tw_{\tau}\cb$.
\eem

{\bf Proof:}
We denote by $\Phi : \ca\ot_{\tau}\cb\ra\cw$ the mapping which is
a left $\ca$--module and a right $\cb$--module isomorphism. We define
$\tau (b\ot a) := \Phi^{-1}(ba)$.
\hfill $\Box$\\

\bem
{\bf Theorem:}
There is one to one correspondence between algebra cross
$\tau :\cb\ot\ca\ra\ca\ot\cb$ and crossed product $\cw$ of
algebras $\ca$ and $\cb$.
\eem

{\bf Proof:} Let us assume that the algebra $\cw$ is universally
generated  crossed product of algebras $\ca$ and $\cb$. We define
a linear mapping $\tau_{\cw} : \cb \ot \ca \ra \ca \ot \cb$ by the
following formula
\be
\tau_{\cw} (b \ot a) :=
[m_{\cw}\circ (\ic \ca, \ot \ic \cb,)]^{-1} (ba).
\ee
It can be deduced that the above mapping is an algebra cross. Moreover,
the map $m_{\cw} \circ (i_{\ca} \ot i_{\cb})$ is an algebra isomorphism
of $\ca \tt \tau, \cb$ onto $\cw$. Conversely, let
$\tau :\cb\ot\ca\ra\ca\ot\cb$ be an algebra cross, then the tensor product
$\ca\ot\cb$ of algebras $\ca$ and $\cb$ equipped with the multiplication
$m_{\tau}: (\ca\ot\cb)\ot(\ca\ot\cb)\ra\ca\ot\cb$ defined by the formula
\be
\ba{c}
m_{\tau} := (m_{\ca} \ot m_{\cb})
\circ (id_{\ca} \ot \tau \ot id_{\cb})
\label{mul}
\ea
\ee
is associative \cite{dk,csv}. In this case
both relations (\ref{mtw}b) and (\ref{mtr}b) can be written equivalently
by
\be
\tau \circ (\m \cb, \ot \m \ca,) =
\m \tau, \circ (\tau \ot \tau)
\circ (id_{\cb} \ot \tau \ot id_{\ca}).
\ee
It is easy to see that $\ca\ot\cb$ equipped with the above
multiplication is a crossed product of algebras $\ca$ and $\cb$.
The inclusion $\ic \ca,$ and $\ic \cb,$ are defined as follows
\be
\ba{c}
\ic \ca, (a) := a \ot 1 ,\;\;\; \ic \cb, (b) := 1 \ot b.
\label{inl}
\ea
\ee
\hfill $\Box$\\

Observe that the multiplication (\ref{mul}) in the crossed product
$\ca \tw_{\tau} \cb$ of algebras $\ca$ and $\cb$ can be given in
the following form
\be
(a' \ot b)(a \ot b') = a'a_{(1)} \ot b_{(2)}b', \ee where, as
already mentioned, the Sweedler type notation for the cross $\tau$
has been used, i.e.
\be
\tau (b \ot a) := a_{(1)} \ot b_{(2)}.
\ee
In particular, we have the relations
\be
(a \ot 1)(1 \ot b) = a \ot b, \;\;\;
(1 \ot b)(a \ot 1) = a_{(1)} \ot b_{(2)}.
\ee
It is interesting that elements of the algebra $\ca\tt \tau, \cb$
can be ordered in such a way that all elements
of the algebra $\cb$  are to the right and elements of the algebra
$\ca$ are to the left. Such ordering is said to be Wick ordering.

As we already mentioned before, the crossed product related to the
standard twist is known as the tensor product of algebras. If we
use the graded twist, then we obtain the graded tensor product of
graded algebras. In the general case, however, an algebra cross is
not so simple.
The construction of all possible crossed products for a given pair
of algebras $\ca, \cb$ is a problem. It is difficult to describe a
general method for an arbitrary pair of algebras. Hence we
restrict our attention to some particular classes of algebras.

Let $\cb$ be a bialgebra. This means that there is an algebra
homomorphism $\D : \cb \ra \cb \ot \cb$ and the counit $\e$.
We have here the following well--known conditions:
the coassociativity
\be
(id \ot \D ) \circ \D = (\D \ot id) \circ \D,
\ee
and two relations for the counit
\be
(\e \ot id) \circ \D = 1 \ot id, \;\;\;
(id \ot \e) \circ \D = id \ot 1.
\ee
An algebra $\ca$ is said to be a left $\cb$--module algebra if
there is an action $\rhd : \cb \ra \ca$ such that
\be
\ba{l} b \rhd (a a') = (b^{(1)} \rhd a)(b^{(2)} \rhd a'),\\ 1 \rhd
a = a . \ea \ee We have the following:

\bem
{\bf Lemma:} If $\ca$ is a left $B$--module algebra, then
there is an algebra cross $\tau : \cb \ot \ca \ra \ca \ot \cb$
defined by the relation
\be
\ba{c} \tau (b \ot a) = (b^{(1)} \rhd a) \ot b^{(2)} \label{twb}
\ea \ee for $a \in \ca$ and $b \in \cb$.\\ \eem {\bf Proof:} We
shall show that two condition (\ref{mtw}, \ref{mtr}) for the cross
(\ref{twb}) are satisfied. For the left hand side of the first
relation (\ref{mtw}) we calculate $$ \ba{l} \left[ \tau \circ
(id_{\cb} \ot m_{\ca}) \right] (b \ot a \ot a')\\ = [b^{(1)} \rhd
m(a \ot a') \ot b^{(2)}]\\ = m \left[ (b^{(1)} \rhd a) \ot
(b^{(2)} \rhd a') \right] \ot b^{(3)} , \ea $$ where $(id \ot \D )
\circ \D (b) = b^{(1)} \ot b^{(2)} \ot b^{(3)}$ and the
coassociativity condition have been used. For right hand side we
obtain $$ \ba{c} \left[ (m_{\ca} \ot id_{\cb}) \circ (id_{\ca} \ot
\tau) \circ (\tau \ot id_{\ca}) \right] (b \ot a \ot a')\\ = m
\left[ (b^{(1)} \rhd a) \ot (b^{(2)} \rhd a') \right] \ot b^{(3)}
. \ea $$ The second relation (\ref{mtr}) can be calculated in a
similar way. \hfill $\Box$\\

If $\tau$ is the cross defined by the formula (\ref{twb}), then the
multiplication in the corresponding crossed product $\ca \tt \tau, \cb$
is given by
\be
(a \ot b)(a' \ot b') = a(b^{(1)} \rhd a') \ot b^{(2)}b'. \ee We
can see, in this case, that the crossed product $\ca \tt \tau,
\cb$ is exactly the so--called smash product $\ca
=\!\!\!\!\!\!/\!\!/ \cb$, see Ref.\cite{cohmont,mon}. If in
addition $\ca$ and $\cb$ are endowed with a Hopf algebra
structure, then the corresponding crossed product is the
semi--simple product of Hopf algebras introduced by Molnar
\cite{mol}.

Let $\ca$ and $\cb$ be a dual pair of Hopf algebras \cite{vd}.
This means that we have a bilinear pairing $<.,.> : \cb \ot \ca
\ra k$ such that
\be
<\D(b), a \ot a'> = <b, aa'>, \;\;\;
<bb', a> = <b \ot b', \D(a)>.
\ee
Observe that there is a left action of the algebra $\cb$ on $\ca$
\be
b \rhd a = <b, a^{(2)}> a^{(1)}, \ee One can prove that the
algebra $\ca$ is a (left) $\cb$--module algebra and the mapping
$\tau : \cb \ot \ca \ra \ca \ot \cb$ defined by
\be
\tau(b \ot a) \equiv a\dn 1, \ot b\dn 2,:= b^{(1)}\rhd a\ot
b^{(2)}=
 <b^{(1)}, a^{(2)}>a^{(1)}\ot b^{(2)}   \ee is a
cross. This is interesting point that the corresponding crossed
product $\ca \tt \tau, \cb$ contains all information about
noncommutative differential operators \cite{spc} on $\ca$. It
means that we can forget the Hopf algebra structures in $\ca$ and
$\cb$ and restrict our attention to the algebra structure only. In
this case we obtain the so--called crossed product of algebras
\cite{csv}.
\section{Free product of algebras}
Let $\ca$ and $\cb$ be two unital associative algebras over a
field $k$. Then there is an algebra of polynomials containing
elements of these two algebras. This algebra is said to be a free
product of algebras $\ca$ and $\cb$ \cite{wang}. Namely, we have
here the

\bem {\bf Definition:} An (algebraic) free product of algebras
$\ca$ and $\cb$ is the algebra $\ca\ast\cb$ formed by all formal
finite sums of monomials of the form $a_1\ast b_1\ast a_2\ast
\ldots$ or $b_1\ast a_1\ast b_2\ast\ldots$, where $a_i \in \ca$,
$b_i \in \cb$, $i = 1, 2,\ldots$ are non-scalar elements. \eem

In other words $\ca \ast \cb$ is the algebra generated by two
algebras $\ca$ and $\cb$ with no relations except for the
identification of unit element, i.e. $1_{\ca} = 1_{\cb} = 1$.
One can see that this free product of algebras is commutative and
associative
\be
\ca\ast \cb = \cb\ast\ca, \;\;\;\; (\ca\ast\cb)\ast\cc =
\ca\ast(\cb\ast\cc). \ee Moreover, if $\ca_1$ is a subalgebra of
$\ca$ and $\cb_1$ is a subalgebra of $\cb$, then $\ca_1 \ast
\cb_1$ is a subalgebra of $\ca \ast \cb$. In particular,  the
algebras $\ca$ and $\cb$ are subalgebras of $\ca\ast\cb$. It is
known that the product $\ca\ast\cb$ possesses the following
universal property:

\bem {\bf Lemma:} For every pair of algebra maps $u : \ca \ra \cc$
and $v : \cb \ra \cc$ there exist one and only one algebra map $w$
such that $u = w \circ \jc \ca,$ and $v = w \circ \jc \cb,$ or in
other words the following diagram
\be
\ba{rllll} &\ca&&&\\ j_{\ca}&\downarrow&&\stackrel{u}{\searrow}&\\
&&&&\\ \ca&\ast&\cb&\stackrel{w}{\ra}&\cc\\ &&&&\\
j_{\cb}&\uparrow&&\stackrel{v}{\nearrow}&\\ &\cb&&& \label{univ}
\ea \ee commutes. Here, $j_{\ca}$ (resp. $j_{\cb}$) denotes the
natural inclusion of $\ca$ (resp. $\cb$) into $\ca\ast\cb$. Note
that $w$ is onto if and only if $\cc$ is generated by images
$u(\ca)$ and $v(\cb)$.\\ \eem Proof: The proof can be immediately
seen if one defines $$ w(\ldots b\ast a\ldots)=\ldots
v(b)u(a)\ldots $$ $\ $ \hfill $\Box$\\

Let us consider a simple example of an algebraic free product.\\
{\bf Example:} Let $U$ and $W$ be two $k$-vector space. Then
the tensor algebra over the direct sum $U \oplus W$ is a free
product of tensors algebras $TU$ and $TW$, i.e. we have the relation
$$
T(U \oplus W) = TU \ast TW.
$$

Let us consider the free product of maps.

\bem
{\bf Definition:}
Let $f : \ca \ra \cb$ and $g : \cc \ra \cd$ be two algebra maps. Then
a mapping $f \ast g : \ca \ast \cb \ra \cc \ast \cd$ defined by
\be
f \ast g\quad (\ldots b \ast a \ldots) = \ldots g(b) \ast f(a) \ldots
\ee
for $a \in \ca$ and $b \in \cb$,
is called a free product of $f$ and $g$.
\eem

We have the following simple lemma.

\bem
{\bf Lemma:}
The map $f \ast g$ is injective (resp. surjective) if and only if
the maps $f$ and $g$ are injective (resp. surjective).\\
\eem
$\ $\hfill $\Box$\\

Now we are going to study ideals in the free product of algebras.
It is interesting that in a free product $\ca \ast \cb$ of
algebras $\ca$ and $\cb$ may exists an ideal $J$ such that the
quotient $(\ca \ast \cb)/J$ can be also expressed as a free
product of certain algebras.

\bem {\bf Lemma:} Let $I_{\ca}$ and $I_{\cb}$ be ideals in
algebras $\ca$ and $\cb$, respectively. Then
\be
J(I_\ca, I_\cb) := I_{\ca} \ast \cb + \ca \ast I_{\cb} \ee forms
an ideal in the free product $\ca \ast \cb$ such that
\be
\ba{c} \ca\ast\cb/J(I_\ca, I_\cb) = \left( \ca/I_{\ca} \right)
\ast \left( \cb/I_{\cb} \right). \ea \ee \eem {\bf Proof:} Let
$\pi_{\ca} : \ca \ra \ca/I_{\ca}$ and $\pi_{\cb} : \cb \ra
\cb/I_{\cb}$ be canonical projections, i.e. $I_\ca:=ker\pi_\ca$
and $I_\cb:=ker\pi_\cb$. It is obvious that the free product
$\pi_{\ca} \ast \pi_{\cb} : \ca \ast \cb \ra (\ca/I_{\ca}) \ast
(\cb/I_{\cb})$ is surjective and $(\ca/I_{\ca}) \ast (\cb/I_{\cb})
= (\ca \ast \cb)/ker(\pi_{\ca} \ast \pi_{\cb})$. One can see that
$ker(\pi_{\ca} \ast \pi_{\cb}) = ker(\pi_{\ca})\ast\cb + \ca\ast
ker(\pi_\cb) = J(I_\ca, I_\cb)$. \hfill $\Box$\\

The ideal $J(I_\ca, I_\cb)$ from the above lemma is called {\it a
free ideal} in $\ca\ast\cb$ generated by $I_{\ca}$ and $I_{\cb}$.

If $\ca$ and $\cb$ are two $k$-algebras and
$\tau : \cb \ot \ca \ra \ca \ot \cb$ is a cross, then
the crossed product $\ca \tw_{\tau} \cb$ of these algebras
can be given by their free product modulo certain ideal.
More precisely, we have here the following

\bem
{\bf Lemma:} For the crossed product $\ca \tw_{\tau} \cb$ of algebras
$\ca$ and $\cb$ we have the formula
\be
\ca \tw_{\tau} \cb = (\ca \ast \cb)/I_{\tau},
\ee
where $I_{\tau}$ is an ideal generated by the relation
\be
I_{\tau} = gen\{b \ast a - a_{(1)} \ast b_{(2)}\} \ee for $a \in
\ca, b \in \cb$ and $\tau (b \ot a) := a_{(1)} \ot b_{(2)}$.\\
\eem {\bf Proof:} We use the universality of the free product. If
$C \equiv \ca \tw_\tau \cb$, then $u \equiv \ic \ca,$, $v \equiv
\ic \cb,$, and there exist unique morphism $w \equiv \ic \ca, \ast
\ic \cb,$ such that $\ic \ca, = w \circ \jc \ca,$ and $\ic \cb, =
w \circ \jc \cb,$. Observe that $w$ is a morphism from $\ca \ast
\cb$ to $\ca \tw_\tau \cb$, and his kernel is equal to the ideal
$I_{\tau}$. \hfill $\Box$\\
\section{Crossed product of free algebras}
Let $\ca$ and $\cb$ are graded algebras. This means that we have the
following decompositions
\be
\ca = \bigoplus\limits_{k=0}^{\infty} \ \ca^k, \;\;\;
\cb = \bigoplus\limits_{k=0}^{\infty} \ \cb^k ,
\ee
where $\ca^0\cong\cb^0\cong k$. Let $\tau : \cb\ot\ca\ra\ca\ot\cb$ be
an arbitrary algebra cross. Then the algebra cross $\tau$ can be given
by the relation
\be
\tau = \bigoplus\limits_{k,l=0}^{\infty} \ \tau_{k,l}, \label{dec}
\ee where $\tau_{k,l}$ is the restriction of the algebra cross
$\tau$ to the space $\cb^k \ot \ca^l$, $(k, l = 1, 2, \ldots)$. In
such a way the algebra cross $\tau$ can be reduced by a set of
mappings $\{\tau_{i,j}:\cb^{i}\ot \ca^{j}\ra \ca\ot\cb\}$. Observe
that we always have
\be
\tau_{0,0} \equiv id,\quad \tau_{k,0}(\cb^{k} \ot 1) := 1\ot
\cb^{k},\quad \tau_{0,m}(1\ot \ca^{m}) := \ca^{m}\ot 1. \ee We
have here the following problem: Find the conditions under which
all $\tau_{k,l}$ for $k, l > 1$ can be constructed starting from
$\tau_{1,1}$. The mapping $\tau_{1,1}$ is given as an initial data
for such construction. We restrict our attention to certain
particular cases. Let us consider the crossed product of free
algebras in details. Let $\ca$ be a free algebra generated by
$x^1,\ldots,x^m$ and let $\cb$ be a free algebra generated by
$y^1,\ldots,y^n$. We identify these free algebras $\ca$, $\cb$
with tensor algebras $TE$ and $TF$, respectively, where $E$ is a
linear span of generators $x^1,\ldots,x^m$ of $\ca$ and $F$ is a
linear span of $y^1,\ldots,y^n$. This means that $\ca^1 \equiv E$,
$\ca^k \equiv E^{\ot k}$, and similarly for $\cb$. Note that $E$
and $F$ are said to be generating spaces for algebras $\ca$ and
$\cb$, respectively. Let us consider the structure of the crossed
product of free algebras $TE$ and $TF$ in more details.

\bem {\bf Remark:} Let $\tau_{1,1} : F\ot E\ra E\ot F$ be a linear
mapping, then there is a unique algebra cross $\tau : TF \ot TE
\ra TE \ot TF$ such that $\tau |_{F\ot E} = \tau_{1,1}$. \eem

Indeed, if $\tau_{1,1} : F\ot E\ra E\ot F$ is a linear mapping,
then we can introduce a set of mappings $\{\tau_{i,j}:F^{\ot i}\ot
E^{\ot j}\ra TE\ot TF\}$ as follows: Obviously $\tau_{0,0} \equiv
id$, $\tau_{k,0}(F^{\ot k} \ot 1) := 1\ot F^{\ot k}$, and
$\tau_{0,m}(1\ot E^{\ot m}) := E^{\ot m}\ot 1$. Then the algebra
cross $\tau$ can be defined by the relations (6) and (33). For
example
\be
\ba{l} \tau_{2,1} = (\tau_{1,1} \ot id) \circ (id \ot \tau_{1,1}).
\ea \ee Similarly
\be
\ba{l} \tau_{1,2} = (id \ot \tau_{1,1}) \circ (\tau_{1,1} \ot id).
\ea \ee We can calculate $\tau_{k,l}$ for arbitrary $k, l$ in a
similar way.
Let us consider this case in more details.\\
\bem {\bf Definition:} Let $\ca$ and $\cb$ be two graded algebras.
An algebra cross $\tau$ is said to be homogeneous
if the image of $\tau_{k,l}$ lies in $\ca^l \ot \cb^k$ for all $k, l
= 1, 2, \ldots$. \eem

It is obvious that he homogeneous cross can be determined uniquely
by a set of linear mappings $\tau_{k,l} : \cb^k \ot \ca^l \ra
\ca^l \ot \cb^k$ such that
\be
\ba{l} \tau_{k,l+m}\circ(id_\ca\ot m_\ca)=(m_\ca\ot
id_\cb)\circ(id_{\ca} \ot \tau_{k,m}) \circ (\tau_{k,l} \ot
id_{\ca}),\\ \tau_{k+l,m}\circ(m_\cb\ot id_\ca) = (id_\ca\ot m_\cb
)\circ (\tau_{k,m} \ot id_{\cb}) \circ (id_{\cb} \ot \tau_{l,m}).
\label{taf} \ea \ee for arbitrary integers $k, l, m > 0$.

Consider two free algebras $\ca := TE$ and $\cb := TF$ with their
natural gradings. Choose a basis $x^1,\ldots,x^m$ in $E$ and a
basis $y^1,\ldots,y^n$ in $F$. Now, the linear operator
$\tau_{1,1} \equiv \hat{\tau} : F \ot E \ra E \ot F$  can be
expressed by
\be
\ba{c} \hat{\tau}(y^i \ot x^j) = \hat{\tau}^{ij}_{kl} \ x^k \ot
y^l , \label{cop} \ea \ee its matrix elements
$\hat{\tau}^{ij}_{kl}$. Let us calculate all components
$\tau_{k,l} : F^{\ot k} \ot E^{\ot l} \ra E^{\ot l} \ot F^{\ot k}$
for this cross. Obviously for $k=1$ and arbitrary $l>1$ we obtain
the map $\tau_{1,l} : F \ot E^{\ot l} \ra E^{\ot l} \ot F$, where
\be
\ba{c} \tau_{1,l} := \hat{\tau}^{(l)}_l \circ \ld \circ
\hat{\tau}^{(1)}_l, \label{ctwi} \ea \ee and $\hat{\tau}^{(i)}_l :
E^{(i)}_l \ra E^{(i+1)}_l$, $E^{(i)}_l := E \ot \ldots \ot E \ot F
\ot E \ot \ldots \ot E$ ($l+1$-factors, $F$ on the i-th place,
$1\leq i \leq l$) is given by the relation $$ \hat{\tau}^{(i)}_l
:= \underbrace{ id_{E} \ot \ldots \ot \hat{\tau} \ot \ldots \ot
id_{E} }_{l\;\;times}, $$ where $\hat{\tau}$ is on the i-th place.
One verifies that $$ \hat\tau^{(i)}_l\circ\hat\tau^{(j)}_l=
\hat\tau^{(j)}_l\circ\hat\tau^{(i)}_l$$ if $|i-j|\ge 2$. For
arbitrary $k\ge 1$ and $l\ge 1$ we obtain the map $\tau_{k,l} :
F^{\ot k} \ot E^{\ot l} \ra E^{\ot l} \ot F^{\ot k}$, where
\be
\ba{l} \tau_{k,l} := (\tau_{1,l})^{(1)}  \circ \ldots \circ
(\tau_{1,l})^{(k)}, \label{up} \ea \ee where $(\tau_{1,l})^{(i)}$
is defined in similar way like $\hat{\tau}_l^{(i)}$. In this way
we obtain the result:

\bem {\bf Lemma:} Let $TE$  and $TF$ be free algebras and
$\hat{\tau}$ be a linear operator defined on generators of these
algebras by the relation (\ref{cop}), then there is a homogeneous
algebra cross $\tau : TF \ot TE \ra TE \ot TF$ which is given by
the relations (\ref{ctwi}) and (\ref{up}). \eem

{\bf Proof:} We must prove that for the map $\tau$ defined by
relations (\ref{ctwi}) and (\ref{up}) the identities (\ref{taf})
hold true. Observe that we have $m_{\ca}(a \ot a^\prime) \equiv a
\ot a^\prime$  for $a\in E^{\ot l}, a'\in E^{\ot m}$ and similarly
for $m_{\cb}$, i.e. $m_\ca$ and $m_\cb$ act as identity operators
in this case. Therefore, the relations (\ref{taf}) can be
rewritten in a simpler form
\be
\ba{l} \tau_{k,l+m}   = (id_{TE} \ot \tau_{k,m}) \circ (\tau_{k,l}
\ot id_{TE}),\\ \tau_{k+l,m}   =  (\tau_{k,m} \ot id_{TF}) \circ
(id_{TF} \ot \tau_{l,m}). \label{crr}\ea \ee
After substituting the definition (\ref{up}) of the maps
$\tau_{k,l}$ into (\ref{crr}) and some calculations we obtain our
result.
\hfill $\Box$\\

We have here the following

\bem {\bf Theorem:} Let $\tau:TF\ot TE\ra TE\ot TF$ be an
arbitrary cross. Then for the corresponding crossed product
we have the following relation
\be
TE \tw_{\tau} TF = T(E \oplus F)/I_{\tau}, \ee where 
\be
I_{\tau} := gen\{b \ot a - a_{(1)} \ot b_{(2)}\} \ee
is an ideal in $T(E \oplus F)$. If the cross $\tau$ is
homogeneous, then
\be
I_{\tau} := gen\{y^i \ot x^j - \hat{\tau}^{ij}_{kl} x^l \ot y^k\}.
\ee \eem
{\bf Proof:} According to the last Lemma of the previous
Section for the crossed product 
we have the relation $$ TE \tw_{\tau} TF = (TE \ast TF)/I_{\tau} =
T(E \oplus F)/I_{\tau}. $$ \hfill $\Box$\\ Now it is obvious that
in the study of noncommutative de Rham complexes and
noncommutative calculi with partial derivatives there are several
examples of algebras which can be described as algebra crossed
product \cite{man2,wam,BK,powo,mukh,dem}.

If the operator $\hat{\tau}$ is given by the diagonal matrix
$\hat{\tau}^{ij}_{kl} := t^{ij} \d^i_l \d^j_k$,
$t^{ij} \in k \setminus \{0\}$, then we obtain a simple
example of cross for free algebras, namely the so called
color cross
\be
\ba{c}
\tau(y^i \ot x^j) = t^{ij} \ x^j \ot y^i,
\ea
\ee
If we assume that $k \equiv \com$ and $t^{ij} \equiv q$,
$q \in \com \setminus \{0\}$, then we obtain the $q$-cross.
\section{Ideals in crossed product}
Let us assume that $\ca \tw_\tau \cb$ and $\ca' \tw_{\tau'} \cb$
are crossed product of algebras $\ca$, $\cb$ and $\ca'$, $\cb'$
with respect to a cross $\tau$ and $\tau'$, respectively. It is
natural to define a morphism of such two crossed products of
algebras as a map which transform the first crossed product in the
second one.

\bem
{\bf Definition:} An algebra morphism
$h : \ca \tt \tau, \cb \ra \ca' \tt \tau', \cb'$ is said to be
a crossed product algebra morphism if there exist two algebra
morphism: $h_{\ca} : \ca \ra \ca'$ and $h_{\cb} : \cb \ra \cb'$
such that $h = h_{\ca} \ot h_{\cb}$.
\eem

The above definition means that the crossed product algebra
morphism $h : \ca \tw_{\tau} \cb \ra \ca' \tw_{\tau'} \cb$ is
described as a pair of algebra homomorphisms $h_{\ca} : \ca \ra
\ca'$ and $h_{\cb} : \cb \ra \cb'$. Observe that in the opposite
case when we have an arbitrary pair of algebra homomorphism, then
their tensor product is not a crossed product algebra morphism,
however, there is the following lemma:

\bem {\bf Lemma:} Let $h_{\ca} : \ca \ra \ca'$ and $h_{\cb} : \cb
\ra \cb'$ be two algebra morphism. Then $h = h_{\ca} \ot h_{\cb}$
is a crossed product algebra morphism if and only if we have the
relation
\be
(h_{\ca} \ot h_{\cb}) \circ \tau =
\tau' \circ (h_{\cb} \ot h_{\ca}),
\ee
or in other words the following diagram commutes
\be
\ba{ccc}
\cb \ot \ca&\stackrel{\tau}{\ra}&\ca \ot \cb\\
_{h_{\cb} \ot h_{\ca}}\downarrow&&\downarrow_{h_{\ca} \ot h_{\cb}}\\
\cb' \ot \ca'&\stackrel{\tau'}{\ra}&\ca' \ot \cb'\\
\label{cotw}
\ea
\ee
\eem
\hfill $\Box$\\

We introduce the notion of ideals in crossed product of algebras.
Let $\ca \tt \tau, \cb$ be a crossed product of algebras $\ca$ and $\cb$
with respect to a cross $\tau$, then we have the following:

\bem {\bf Definition:} A two--sided ideal $J$ in $\ca \tt \tau,
\cb$ is said to be a crossed ideal in $\ca \tt \tau, \cb$ if the
quotient map $\pi : \ca \tt \tau, \cb  \ra (\ca \tt \tau, \cb)/J$
is a morphism of crossed products of algebras. \eem

The above definition means that the factor algebra $(\ca \tw_\tau
\cb)/J$, where $J$ is a crossed ideal must be a crossed product of
certain algebras $\ca'$, $\cb'$ with respect to a certain new
cross $\tau'$. Thus we must have the relation
\be
\ba{c} (\ca \tt \tau, \cb)/J \cong \ca' \tt \tau', \cb'.
                                                      \label{idem}
\ea \ee Let us consider this problem in more details. If $\pi :
\ca \tt \tau, \cb \ra (\ca \tt \tau, \cb)/J$ is a surjective
morphism of a crossed product of algebras, then there is a pair of
surjective algebra homomorphisms $\pi_{\ca} : \ca \ra \ca'$ and
$\pi_{\cb} : \cb \ra \cb'$. Observe that these mappings are in
fact quotient ones. This means that $\ca' \equiv \ca/I_{\ca}$, and
$\cb' \equiv \cb/I_{\cb}$ where $I_{\ca}$ (the kernel of
$\pi_\ca$) is a two-sided ideal in $\ca$ and $I_{\cb}$ is a
two-sided one in $\cb$. One can see that there is a cross $\tau' :
\cb/I_{\cb} \ot \ca/I_{\ca} \ra \ca/I_{\ca} \ot \cb/I_{\cb}$ such
that the following diagram is commutative
\be
\ba{ccc}
\cb \ot \ca&\stackrel{\tau}{\ra}&\ca \ot \cb\\
_{\pi_{\cb} \ot \pi_{\ca}}\downarrow&&\downarrow_{\pi_{\ca}
\ot \pi_{\cb}}\\
\cb/I_{\cb} \ot \ca/I_{\ca}&\stackrel{\tau'}{\ra}&\ca/I_{\ca}
\ot \cb/I_{\cb}.
\ea
\ee
In this way we obtain the following:

\bem {\bf Lemma:} If $J$ is a crossed ideal in the crossed product
$\ca \tw_\tau \cb$, then there is a pair of ideals $(I_{\ca},
I_{\cb})$ in algebras $\ca$ and $\cb$, respectively and the cross
$\tau' : \cb/I_{\cb} \ot \ca/I_{\ca} \ra \ca/I_{\ca} \ot
\cb/I_{\cb}$ such that we have the relation (\ref{idem}). \eem
\hfill $\Box$\\

It is interesting to investigate the opposite statement. For a
given ideals $(I_{\ca}, I_{\cb})$ in $\ca$ and $\cb$, respectively
find a corresponding ideal in $\ca \tt \tau, \cb$. First, we
consider a particular case when one of the ideal in the above pair
is trivial.

\bem {\bf Definition:} A two--sided ideal $I_{\ca}$ in the algebra
$\ca$ such that  $I_{\ca}\ot\cb$ is a crossed ideal in the algebra
$\cw \equiv \ca \tw_\tau \cb$ is said to be a left $\tau$-ideal in
$\ca$. \eem

Observe that we have the following criterion (cf. Proposition
3.2.4 in \cite{BKO3})

\bem {\bf Lemma:} An ideal $I_{\ca}$ in $\ca$ is a left
$\tau$-ideal in $\ca \tt \tau, \cb$ if and only if
\be
\tau (\cb \ot I_{\ca}) \subset I_{\ca} \ot \cb .\label{twco} \ee
\eem

{\bf Proof:} How can be easily seen, the condition (\ref{twco}) is
equivalent to the fact that  $J:= I_{\ca}\ot\cb$ is a two-sided
ideal in $\ca \tt \tau, \cb$. Therefore, one has to prove that
(\ref{twco}) implies that $J$ is a crossed ideal as well. Indeed:
the vector space quotient $(\ca\ot\cb)/J$ is isomorphic to
$\ca/I_\ca\ot\cb$. Since $J$ is an ideal, the projection map
$\pi_\ca\ot id_\cb :\ca \tt \tau, \cb\ra \ca/I_\ca\ot \cb$, where
$\pi_\ca (a)=[a]$ and $[a]\in\ca/I_\ca$ denotes the equivalence
class of $a\in\ca$, is an algebra map. In particular, $\pi_\ca\ot
id_\cb ((1\ot b)(a\ot 1))=([1]\ot b)\ot ([a]\ot 1)=[a_{(1)}]\ot
b_{(2)}$. This means that $\tau'([a]\ot b):=[a_{(1)}]\ot b_{(2)}$
is a new twist converting $\ca/I_\ca\ot\cb$ into a crossed product
algebra. Moreover, the following diagram
\be
\ba{ccc} \cb \ot \ca&\stackrel{\tau}{\ra}&\ca \ot \cb\\ _{id_{\cb}
\ot \pi_{\ca}}\downarrow&&\downarrow_{\pi_{\ca} \ot id_{\cb}}\\
\cb \ot \ca/I_{\ca}&\stackrel{\tau'}{\ra}&\ca/I_{\ca} \ot \cb \ea
. \ee must commutes. The formula (\ref{twco}) gives the condition
for the commutativity of the above diagram. \hfill $\Box$\\

It follows immediately from the proof of the previous lemma that
we have: \bem {\bf Lemma:} If $I_{\ca}$ is a left $\tau$-ideal in
the algebra $\ca$, then there is a new cross $\tau' : \cb \ot
\ca/I_{\ca} \ra \ca/I_{\ca} \ot \cb$ such that the quotient
algebra $(\ca\tw_\tau\cb)/(I_{\ca}\ot \cb)$ is isomorphic to the
crossed product $\ca/I_{\ca} \tw_{\tau'} \cb$. \eem \hfill
$\Box$\\

This lemma means that for a left $\tau$--ideal $I_{\ca}$ in $\ca$
we have the relation
\be
(\ca\tt\tau, \cb) /(I_{\ca}\ot \cb) \cong \ca/I_{\ca} \tt \tau',
\cb. \ee This algebra  is said to be a left factor of the crossed
product $\ca \tt \tau, \cb$.

We can define a right $\tau$-ideal $I_{\cb}$ in $\cb$ in a similar
way. It is easy to see that for this ideal we have similar results
as for the left $\tau$-ideal. In this way we obtain a right factor
of the crossed product $\ca \tt \tau, \cb$ as the following
quotient
\be
(\ca \tt \tau, \cb)/(\ca\ot I_{\cb}) \cong \ca \tw_{\tau'}
\cb/I_{\cb} . \ee
 A two-sided ideal $J_{I_\ca , I_\cb} := I_{\ca}\ot\cb +
 \ca\ot I_\cb$
in $\ca \tt \tau, \cb$, where $I_{\ca}$ is a left $\tau$-ideal in
$\ca$ and $I_{\cb}$ is a right $\tau$-ideal in $\cb$, is said to
be a crossed ideal generated by $I_\ca, I_\cb$.

\bem {\bf Theorem:} If $J_{I_\ca, I_\cb}$ is a crossed ideal in
the algebra $\ca\tt \tau, \cb$ generated by $\tau$--ideals $I_\ca,
I_\cb$, then there is a new cross $\tau' : \cb/I_{\cb} \ot
\ca/I_{\ca} \ra \ca/I_{\ca} \ot \cb/I_{\cb}$ such that the
quotient algebra $(\ca \tt \tau, \cb)/J_{I_\ca, I_\cb}$ is
isomorphic to the crossed product of $\ca/I_{\ca}$ and
$\cb/I_{\cb}$. \eem \hfill $\Box$\\

The quotient algebra $(\ca \tt \tau, \cb)/J_{I_\ca, I_\cb}$ is
said to be a factor of the  crossed product $\ca\tw_\tau \cb$ with
respect to $\tau$--ideals $(I_{\ca}, I_{\cb})$.\\

\bem {\bf Definition:} Let $\ca \tt \tau, \cb$ be a crossed
product of algebras $\ca$ and $\cb$ with respect to a given cross
$\tau : \cb \ot \ca \ra \ca \ot \cb$. If there exist a pair of
algebras $\tilde{\ca}$, $\tilde{\cb}$ and a cross $\tilde{\tau} :
\tilde{\cb} \ot \tilde{\ca} \ra \tilde{\ca} \ot \tilde{\cb}$ such
that the product $\ca \tt \tau, \cb$ is an image of $\tilde{\ca}
\ot_{\tilde{\tau}} \tilde{\cb}$ under certain surjective  morphism
$h = (h_{\ca}, h_{\cb})$ of crossed products, i.e. the following
diagram
\be
\ba{ccc}
\tilde{\cb} \ot \tilde{\ca}&\stackrel{\tilde{\tau}}{\ra}
&\tilde{\ca} \ot \tilde{\cb}\\
_{h_{\cb} \ot h_{\ca}}\downarrow&&\downarrow_{h_{\ca} \ot h_{\cb}}\\
\cb \ot \ca&\stackrel{\tau}{\ra}&\ca \ot \cb\\
\ea
\ee
is commutative, then $\tilde{\ca} \ot_{\tilde{\tau}} \tilde{\cb}$
is said to be a cover crossed product for $\ca \tw_{\tau} \cb$.\\
\eem

\bem {\bf Lemma:} Assume that $\ca$ and $\cb$ are algebras with
presentation $\ca := TE/I_{\ca}$ and $\cb := TF/I_{\cb}$. If
$\tilde{\tau} : TF \ot TE \ra TE \ot TF$ is a cross, then the
corresponding crossed product $TE \tw_{\tilde{\tau}} TF$ is cover
for a product $\ca \tt \tau, \cb$ with certain cross $\tau : \cb
\ot \ca \ra \ca \ot \cb$ if and only if the ideal $I_{\ca}$ is a
left $\tilde{\tau}$-ideal in $TE$ and $I_{\cb}$ is a right
$\tilde{\tau}$-ideal in $TF$. \eem \hfill $\Box$\\

\bem {\bf Lemma:} Let $\ca \tt \tau, \cb$ be a crossed product of
algebras $\ca$ and $\cb$ with presentation $\ca := TE/I_{\ca}$ and
$\cb := TF/I_{\cb}$, respectively. If the crossed product $TE
\tw_{\tilde{\tau}} TF$ is a cover for the product $\ca \tt \tau,
\cb$, then we have the relation
\be
\ba{c}
\ca \tt \tau, \cb \equiv T(E \oplus F)/I,
\ea
\ee
where $I$ is the ideal in the tensor algebra $T(E \oplus F)$ of the form
\be
I \equiv I_1 + I_2 + I_{\tilde\tau}, \ee $I_1 := \lb
I_{\ca}\rb_{T(E\oplus F)}$ is an ideal in $T(E\oplus F)$ generated
by the $\tilde\tau$--ideal $I_{\ca}$, similarly $I_2 := \lb
I_{\cb}\rb_{T(E\oplus F)}$, and $I_{\tilde \tau}$ is an ideal in
$T(E \oplus F)$ defined by the relation
\be
I_{\tilde \tau} := \lb v \ot u - \tilde \tau(v \ot
u)\rb_{T(E\oplus F)}, \ee for every $u \in E$ and $v \in F$. \eem
\hfill $\Box$\\

\bem {\bf Lemma:} Assume that $E, F$ are two linear spaces and
$R:E\ot E\ra E\ot E$, $S: F\ot F\ra F\ot F$
are two linear operators. Let $\ca$ and $\cb$ be two quadratic
algebras generated by $E$ and $F$. It means that we have the
quotients
\be
\ca := TE/I_{R},\;\;\; \cb := TF/I_{S}, \ee where ideals  are
given by the quadratic relations $$ I_{R} = \lb id - R\rb_{TE},
\;\;\; I_{S} = \lb id - S\rb_{TF}. $$ Assume further, that  a
homogeneous cross $\tilde\tau: TF\ot TE\ra TE\ot TF$ is induced by
a linear operator $C : F \ot E \ra E \ot F$.

Then there is a cross $\tau : \cb \ot \ca \ra \ca \ot \cb$ and the
corresponding crossed product $\ca \tt \tau, \cb$ if and only if
we have the following relations
\be
\ba{l} (id \ot C) \circ (C \ot id) \circ (id - (id \ot R)) = (id -
(R \ot id)) \circ (id \ot C) \circ (C \ot id),\\ (C \ot id) \circ
(id \ot C) \circ (id - (S \ot id)) = (id - (id \ot S)) \circ (C
\ot id) \circ (id \ot C). \label{coco} \ea \ee
Moreover
\be
T(E \oplus F)/I \cong TE/I_{R} \tt \tau, TF/I_{S}, \ee where $I$
is an ideal of the form
\be
I \equiv I_1 + I_2 + I_{C}, \ee $I_1 := \lb I_{R}\rb_{T(E\oplus
F)}$, $I_2 := \lb I_{S}\rb_{T(E\oplus F)}$ and $I_{C}$ is an ideal
given by the relation
\be
I_{C} := \lb v \ot u - C(v \ot u)\rb_{T(E\oplus F)}, \ee for every
$u \in E$ and $v \in F$. \eem

{\bf Proof:} One checks that (\ref{coco}) are equivalent to the
$\tilde\tau$-ideal conditions (\ref{twco}) for $I_R$ and $I_S$
respectively.
\hfill $\Box$\\

\bem
{\bf Lemma:} Let $R$, $C$ and $S$ be three linear operators
like in the previous lemma. If
\be
\ba{l} (R \ot id) (id \ot C) (C \ot id) = (id \ot C) (C \ot id)
(id \ot R),\\ (id \ot S) (C \ot id) (id \ot C) = (C \ot id) (id
\ot C) (S \ot id), \ea \ee then the conditions (\ref{coco}) are
satisfied.
 \eem \hfill $\Box$\\ Let us consider an example of an algebra
crossed product. It is well--known that the noncommutative
differential calculi with partial derivatives on the quantum plane
is determined by an $R$--matrix satisfying the following braid
relation
\be
R^{(1)} R^{(2)} R^{(1)} = R^{(2)} R^{(1)} R^{(2)},
\ee
and the Hecke condition
\be
(R - q)(R + q^{-1}) = 0, \ee where $q \in \com \setminus \{0\}$.
Here $R^1$ means $R\ot{\rm id}_E$ and analogously $R^2:={\rm
id}_E\ot R$. The noncommutative coordinate algebra and the
corresponding partial derivatives algebra can be expressed as the
following quotients $\ca := TE/I_{\ca}$ and $\cb := TE'/I_{\cb}$,
where ideals $I_{\ca}$ and $I_{\cb}$ are generated by the
quadratic relations $$ I_{\ca} := \lb id - q^{-1} R\rb_{TE},
\;\;\; I_{\cb} := \lb id - q^{-1} R^t\rb_{TE'},$$ $R^t$ is the
transpose of $R$ and $E'$ is the algebraic dual of $E$. It follows
from the previous lemma and Wess--Zumino consistency conditions
\cite{WZ}, that there is a cross $\tau_R$ and the corresponding
crossed product $\cw (R) := \ca \tt {\tau_R}, \cb$. In this case
we must replace $R$ by $q^{-1}R$, $S := q^{-1}R^t$, and $C := qR$.
The Hecke relation solves a linear Wess-Zumino consistency
condition (see also \cite{BKO2,BK,BKO3} in this context). Observe
that $\cw (R)$ is just the quantum Weyl algebra considered
previously by Giaquinto and Zhang \cite{gzh}.

As we already mentioned, $(\ca,\cb)$--modules can be identify with
left $\ca\ot_c\cb^\op$--modules. Thus left $\ca \tw_{\tau}
\cb^\op$--modules supply a generalization for the
$(\ca,\cb)$--bimodule structure (2). Below, by using the Fock
space representation methods (the quantization), we are going
to describe some class of $\ca \tw_{\tau} \ca^\op$--modules.
\section{Crossed envelopings and representations}
The notion of $\ast$--algebras is well--known.
An associative algebra $\ca$ is said to be a $\ast$--algebra if
we have the following relations for the $\ast$--operation
\be
(a b)^{\ast} = b^{\ast} a^{\ast}, \quad a^{\ast\ast} = a , \quad
(\a a)^{\ast} = \overline{\a} a^{\ast}, \ee where $a, b, a^{\ast},
b^{\ast}\in\ca$, $\a\in\com$, $\overline{\a}$ is a complex
conjugated to $\a$. In this section we assume that $k\equiv\com$
is the field of complex numbers. It is obvious that not every
algebra is a $\ast$--algebra. Observe that if $\ca$ is a
$\ast$--algebra, then the $\ast$--operation can be described in
two equivalent ways: as an involutive anti-isomorphism of $\ca$ or
an involutive isomorphism between $\ca$ and $\overline{\ca}^{op}$.
If $\ca$ is not a $\ast$--algebra, then it is interesting to
describe all possible $\ast$--algebra extensions of it. We
introduce here the concept of conjugated algebras and crossed
enveloping algebras for this goal.

\bem {\bf Definition:} If $\ca$ be an arbitrary associative
algebra, then an algebra $\cb$ is said to be conjugated to $\ca$
if there is an antilinear anti-isomorphism (in the complex case)
$(-)^{\star} : \ca \ra \cb$ such that
\be
(a b)^{\star} = b^{\star} a^{\star}, \quad (\a a)^{\star} =
\overline{\a} a^{\star}, \ee where $a, b \in \ca$ and $a^{\star},
b^{\star}$ are their images under the isomorphism $(-)^{\star}$.
\eem

The inverse isomorphism $\cb\ra\ca$ will be denoted by the
same symbol, i.e.
\be
(a^{\star})^{\star} = a.
\ee
If $\ca$ is an algebra, then the conjugated algebra will be denoted
by $\ca^{\star}$. It follows immediately from the definition that for
a given algebra $\ca$ the conjugate algebra $\ca^{\star}$ always exists.
The algebra $\ca^{\star}$ as a vector space is isomorphic to the complex
conjugate space $\overline{\ca}$, and as an algebra -- to the opposite
one $\ca^{op}$, i. e. $\ca \equiv \overline{\ca}^{op}$,
(in the real case it coincides with the opposite algebra $\ca^{op}$).

Consider a crossed product $\cw_{\tau}(\ca):=\ca \tw_{\tau}
\ca^{\star}$ of an algebra with its conjugate.
 We can try to define
the natural $\ast$-operation in $\cw_{\tau}(\ca)$ by the relation
\be
(a \ot b^\star)^{\ast} := b \ot a^{\star} \ee for $a, b \in \ca$.
Then the following holds

\bem {\bf Lemma:} The algebra $\cw_{\tau}(\ca)$ is a
$\ast$-algebra if and only if
\be
\ba{c} (\tau (b^{\star} \ot a))^{\ast} = \tau (a^{\star} \ot b)
\label{scr} \ea \ee for any $a, b\in\ca$. \eem

{\bf Proof:} One needs the property $[(1\ot b^\star)(a\ot 1)]^\ast
= (1\ot a^\star)(b\ot 1)$ or $$b_{(2)}\ot a^\star_{(1)}=b_{(1)}\ot
a^\star_{(2)} $$
which is equivalent to the the relation (\ref{scr}).

\hfill $\Box$\\ An algebra cross $\tau : \ca^{\star} \ot \ca \ra
\ca \ot \ca^{\star}$ satisfying the relation (\ref{scr}) is called
a $\star$--cross.

\bem {\bf Definition:} If
$\tau : \ca^{\star}\ot\ca\ra\ca\ot\ca^{\star}$ is a $\star$--
cross, then the crossed product
\be
\ba{c} \cw_{\tau}(\ca) := \ca \tw_{\tau} \ca^{\star}. \label{wck}
\ea \ee is called a crossed enveloping algebra of $\ca$ with
respect to $\tau$. \eem

Note that the switch $\sigma (b^\star\ot a):=a\ot b^\star$
satisfies (\ref{scr}). It implies that crossed enveloping algebra
generalizes the concept of enveloping algebras for associative
algebra \cite{pie}.

From now on, we assume that every algebra cross considered below
is a $\star$--cross. We shall also identify our two
"star"--operations and use the symbol $"\star"$ for both of them.

Let us consider the crossed enveloping algebra $\cw_{\tau}(\ca)$,
where $\ca\equiv TE$ is a free algebra and the generating space
$E$ is a (finite or infinite dimensional) complex Hilbert space
equipped with an orthonormal
basis $\{x^i : i=1,\cdots ,N\}$, and $\tau$ is an arbitrary cross.
Note that similar algebras have been studied previously by a few
authors \cite{das,jws}. Observe that the conjugated algebra
$\ca^{\star}$ can be identified with the tensor algebra
$TE^{\star}$, where $E^{\star}$ is the complex conjugation space.
The pairing $\pr : E^{\star} \ot E \ra \com$ and the corresponding
scalar product is given by
\be
g_E (x^{\star i}\ot x^j) \equiv (x^{\star i}|x^j)
= \lb x^{i}|x^j \rb := \d^{ij}.
\ee

Let $\hat{\tau} : E^{\star}\ot E\ra E \ot E^{\star}$ be a linear
and Hermitian operator with matrix elements
\be
\ba{c}
\hat{\tau}(x^{\star i}\ot x^j) = \Sigma \ \hat{\tau}^{ij}_{kl}
x^k \ot x^{\star l},
\label{cross}
\ea
\ee
then the quotient
\be
\cw(\hat\tau) = T(E \oplus E^{\star})/I_{\hat\tau}, \ee
where the ideal $I_{\hat\tau}$ is given by the relation
\be
I_{\hat\tau} := gen\{x^{\star i} \ot x^j - \Sigma \
\hat{\tau}^{ij}_{kl} x^k \ot x^{\star l} - (x^{\star i}|x^j)\} \ee
is said to be Hermitian Wick algebra \cite{js}.

\bem {\bf Theorem.}( J\o rgensen,  Schmitt and  Werner \cite{js})
The Hermitian Wick algebra $\cw(\hat\tau)$ is isomorphic to
the crossed enveloping algebra $\cw_{\tau}(TE)$ of $TE$
with respect to the (non-homogeneous) cross generated by
$\hat{\tau} + g_E$. \eem

{\bf Proof:} It has been shownn in \cite{js} that the Wick ordered
monomials form a basis in $\cw(\hat\tau)$. In our language it
means that $\cw(\hat\tau)$ as a vector space is isomorphic to
$TE\ot TE^\ast$. Moreover, $TE$ and $TE^\ast$ are subalgebras in
$\cw(\hat\tau)$. This implies that $\cw(\hat\tau)$ is a crossed
product, i.e. $\cw(\hat\tau)\cong TE\tw_\tau TE^\ast$ for a
certain cross $\tau$. Since $\hat\tau$ is Hermitian operator and
$\langle\,|\,\rangle$ is Hermitian scalar product then
$I_{\hat\tau}$ is $\star$--ideal (i.e. $I_{\hat\tau}^\ast\subset
I_{\hat\tau}$). As a consequence, the cross $\tau$ is
$\star$--cross. \hfill $\Box$\\

Let $\ca$ be an algebra with the presentation $\ca := TE/I_\ca$,
then for the algebra $\ca^{\star}$ we have the presentation
$\ca^{\star} := TE^{\star}/I^{\star}_\ca$. Observe that if $I_\ca$
is a left $\tau$--ideal then $I^{\star}_\ca$ is automatically a
right $\tau$--ideal (remember that $\tau$ is a $\star$--cross).
Then there is a cross $\tau' :
\ca^{\star}\ot\ca\ra\ca\ot\ca^{\star}$ and the corresponding
crossed enveloping algebra $\cw_{\tau'} (\ca) = \ca\tt\tau',
\ca^{\star}$.

Let $H$ be a $k$-vector space. We denote by $L(H)$ the algebra
of linear operators acting on $H$. Let $\ca$ and $\cb$ be two arbitrary
$k$-algebras and $\tau : \cb \ot \ca \ra \ca \ot \cb$ be a cross.\\

\bem {\bf Theorem:}  Let $\pi_{\ca}$ and $\pi_{\cb}$ be
representations of the algebras $\ca$ and $\cb$ in $L(H)$,
respectively. If the condition
\be
\ba{c} \pi_{\cb}(b) \pi_{\ca}(a) = \pi_{\ca}(a_{(1)})
\pi_{\cb}(b_{(2)}) \label{cre} \ea \ee holds for all $a \in \ca$,
$b \in \cb$ , then there exist unique representation $\pi$ of the
crossed product $\ca \tw_{\tau} \cb$ in $L(H)$ such that $\pi
|_{\ca} = \pi_{\ca}$ and $\pi |_{\cb} = \pi_{\cb}$. \eem

{\bf Proof:} The representation $\pi: \ca \tw_{\tau} \cb \ra L(H)$
is defined by
\be
\pi(a \ot b) := \pi_{\ca}(a) \pi_{\cb}(b).
\ee
\hfill $\Box$\\

This theorem allows us to introduce the following definition:\\

\bem
{\bf Definition:}
The representation $\pi$ from the above theorem is said to be
a crossed product of representations $\pi_{\ca}$ and $\pi_{\cb}$
and it is denoted by $\pi_{\ca} \tw_{\tau} \pi_{\cb}$.\\
\eem

It is not difficult to prove the converse:\\

\bem {\bf Theorem:} If $\pi$ is a representation of the crossed
product $\ca \tt \tau, \cb$ of algebras $\ca$ and $\cb$ in $H$,
then there exist representations $\pi_{\ca}$ and $\pi_{\cb}$ of
$\ca$ and $\cb$, respectively, such that
\be
\pi = \pi_{\ca} \tt \tau, \pi_{\cb} .
\ee
\eem
{\bf Proof:} Representations $\pi_{\ca}$ and $\pi_{\cb}$ are defined by
the formulae
\be
\pi_{\ca}(a) := \pi (a\ot 1), \quad
\pi_{\cb}(b) := \pi (1\ot b).
\ee
\hfill $\Box$\\

Let us consider representations of crossed enveloping algebras.
For a given representation $\pi:\ca\ra L(H)$ of $\ca$ in a Hilbert
space $H$ one can define a conjugate representation $\pi_+
(a^\star):=\pi (a)^+$ of the algebra $\ca^\star$, where $+$ stands
for the Hermitian conjugation in $L(H)$.  Thus we have:

 \bem {\bf
Theorem:} Let $\cw \equiv \ca \tt \tau, \ca^{\star}$ be a crossed
enveloping algebra. If $\pi: \ca \ra L(H)$ is a representation in
a Hilbert space $H$, such that 
\be
\ba{c} \pi (b)^+ \pi (a) = \pi (a_{(1)}) \pi (b_{(2)})^{+}
\label{wre} \ea \ee then there is a unique Hermitian or $\star$--representation
$\pi_{\cw} : \cw \ra L(H)$ such that \be \ba{c}\pi_\cw=\pi\tt
\tau,\pi_+ \label{wre1} \ea \ee Conversely, any
Hermitian representation of $\pi_{\cw}$ in a Hilbert space has the
form (\ref{wre1}). \eem

{\bf Proof:} It is a direct consequence of two proceeding
Theorems. One also easily verifies the Hermiticity condition: $\pi_\cw
(w^\star)=\pi_\cw (w)^+$ for $w\in\cw$.
\hfill $\Box$\\

As an example, we outline  the Fock space representation
construction of a cross enveloping algebra $\cw_\tau(\ca)
\equiv \ca \tt \tau,\ca^{\star}$.
For this purpose we assume that $\ca$ is a pre-Hilbert space with
an unitary scalar product $\langle\,|\,\rangle$. Its completion
will be denoted by $H$. In this case we have at our disposal a
canonical representation ({\it the quantization}) $\Pi$ acting on
the algebra $\ca$ by means of the left (or right) multiplications
in $\ca$. For $x,f\in\ca$ it writes
\be
\ba{c} \Pi(x)f:=xf\label{q1} \ea \ee The operators $\Pi (x)$ are,
in general, unbounded operators in $H$. Thus,
\be
\langle\Pi_{+}(x^{\star})f|g\rangle \equiv
\langle\Pi(x)^+f|g\rangle = \langle f|\Pi(x)g\rangle . \ee
 Note that the relations (\ref{wre}) are said to be a {\it commutation
relation} if they are satisfied  for a given (Hermitian) scalar
product $\langle\,|\,\rangle$ on $\ca$. A proper definition of the
action of the operators $\Pi(x)^+$ on the whole algebra $\ca$ can
be a problem. In a case it can be solved, it leads to the canonical
(Fock type) Hermitian representation $\Pi_\cw\equiv\Pi\tw_\tau\Pi_+$ of
the cross enveloping algebra $\cw_\tau(\ca)$ on $\ca$.

If the algebra $\ca$ has generators $\{x^i\}_{i=1,\ldots,N}$,  we
use the notation
\be
\Pi(x^i) \equiv a_{i}^+ , \quad \Pi_{+}(x^{i\star}) \equiv a_{i} ,
\ee It is customary to call them creation and annihilation operators.
Thus the commutation relations play a role of the compatibility
conditions relating $a^+_i$, $\tau$ and $\langle\,|\,\rangle$,
since $a_i$ are Hermitian conjugate to $a^+_i$.  For non-free
algebra $a^+_i$-s have to satisfy a set of generating relations of
the algebra $\ca$. These give rise to the supplementary
commutation relations.

For the ground state $|0\rangle\equiv 1\in\ca$ and annihilation
operators we usually assume
\be
\langle 0|0 \rangle = 0, \quad a_{i} |0\rangle = |0\rangle\
.\label{q2}
\ee

If the action $\Pi$ admits  non-degenerate, positive definite
(pre-) Hermitian scalar product such that the annihilation
operators are well defined and the creation and annihilation
operators satisfy the commutation relations (\ref{wre}) together
with (\ref{q2}),
then we say that we have the well--defined Fock representation for
a crossed enveloping algebra. Of course, the canonical commutation
relations (CCR) and the canonical anti-commutation relations (CAR)
provide the most familiar examples of this type. Some other examples can
be found in \cite{js} and references therein.
Similarly,  systems with generalized statistics can be described
as Fock-like representations of crossed enveloping algebras
\cite{qstat}.
\section*{Acknowledgments}
The authors are  gratefully indebted to Shahn Majid  for several
helpful comments. Some work on this paper was done during the
second author's visit to the University of Kaiserslautern . That
visit was financed by the DAAD. This research was also supported
by KBN under grants 2 P03B 109 15 and 2 P03B 130 12.  One of as
(AB) was also partially supported by Mexican CONACyT \# 27670E and
UNAM -- DGAP \# IN-109599. Finally, we would like to thank the
Referee for numerous improvements and suggestions.


\begin{thebibliography}{99}
\bibitem{mon} S. Montgomery, Hopf algebras and their actions on rings,
Regional Conference series in Mathematics, No 82, AMS (1993).
\bibitem{sm} S. Majid, Int. J. Mod. Phys. {\bf A5}, 1 (1990).
\bibitem{sm2} S. Majid, J. Algebra {\bf 130}, 17 (1990).
\bibitem{sm3} S. Majid, Lett. Math. Phys. {\bf 22}, 167 (1991).
\bibitem{sm4} S. Majid, J. Algebra {\bf 163}, 191 (1994).
\bibitem{dk} A. Van Daele and S. Van Keer, Compositio
Mathematica {\bf 91}, 201 (1994).
\bibitem{csv} A. \v Cap, H. Schichl, J. Van\v zura,
On twisted tensor product of algebras, Comm. Algebra {\bf 23},
4701 (1995).
\bibitem{wor} S. I. Woronowicz, {\em An example of a braided
locally compact group}, in Proceedings of the IXth Max Born Symposium,
Karpacz, September 25 - September 28, 1996, Poland ed by J. Lukierski
et al., Polish Scientific Publishers, Warszawa 1997.
\bibitem{zak} S. Zakrzewski, J. Phys. {\bf A34}, 2929 (1998).
\bibitem{man2}
Yu. I. Manin, Notes on Qauntum Groups and Quantum De Rham
Complexes, Theoret. Math. Phys. {\bf 92}, 997 (1992).
\bibitem{sud}
A. Sudbery, Matrix--element bialgebras determind by quadratic
coordinate algebras, J. Algebra {\bf 158} 375-399 (1993).
\bibitem{wam}
M. Wambst, Ann. Inst. Fourier, Grenoble {\bf 43}, 1089 (1993).
\bibitem{bor}
A. Borowiec, Vector fields and differential operators:
noncommutative case, Czech. J. of Physics {\bf 47}, 1093 (1997)
(q-alg/9710006).
\bibitem{BKO2}
A. Borowiec and V. K.  Kharchenko,  First order optimum calculi,
 Bull. Soc. Sci. Lett. {\L}\'od\'z {\bf 45}, Ser. Recher. Deform.
 XIX, 75 (1995) (q-alg/9501024).
\bibitem{BK}
A. Borowiec and V. K.  Kharchenko,  Differential calculi with
partial derivatives, Siberian Advances in  Mathematics {\bf 5}, 10
(1995).
\bibitem{BKO3}
A. Borowiec, V. K. Kharchenko and Z. Oziewicz,  First order
calculi with values in right--universal bimodules, in: Quantum
Groups and Quantum Spaces, R. Budzy\'nski, W. Pusz and S.
Zakrzewski (Eds.), Banach Center Publications Vol. {\bf 40}, 171
Warszawa (1997) ( q-alg/9609010).
\bibitem{js}
P.E.T. J\o rgensen, L.M. Schmitt, and R.F. Werner, Positive
representation of general commutation relations allowing Wick
ordering, J. Funct. Anal. {\bf 134}, 33 (1995).
\bibitem{mar3}
W. Marcinek, Categories and quantum statistics,
Rep. Math. Phys. {\bf 38}, 149-179 (1996)
\bibitem{mar4}
W. Marcinek, On quantum Weyl algebras and generalized quons, in:
Quantum Groups and Quantum Spaces, R. Budzy\'nski, W. Pusz and S.
Zakrzewski (Eds.), Banach Center Publications Vol. {\bf 40}, 397
Warszawa (1997).
\bibitem{qstat}
W. Marcinek, Remarks on Quantum Statistics, in:Particles, Fields
and Gravitation, J. Rembieli\' nski (Ed.), CP453, American
Institute of Physics, NY (1998) (math.QA./9806158.)
\bibitem{mol}
R. Molnar, J. Algebra {\bf 47}, 29 (1977).
\bibitem{cohmont}
M. Cohen and S. Montogomery, Trans. Am. Math. Soc. {\bf 282},
237 (1984).
\bibitem{wang} S. Wang,
Free product of compact quantum groups, Commun. Math. Phys. {\bf
167}, 671 (1995)
\bibitem{spc}
P. Schupp, (1996), {\em Cartatn calculus: differential geometry
for quantum groups}, in {\em Quantum groups and their applications
in physics} (Varenna 1994), Proc. Internat. School Phys. Enrico
Fermi, 127, IOS, Amsterdam (1996).
\bibitem{vd}
A. Van Daele, Dual pairs of $*$-Hopf algebras, Bull. London Math.
Soc. {\bf 25}, 209 (1993).
\bibitem{nil}
F. Nill, Rev. Math. Phys. {\bf 6}, 149 (1994).
\bibitem{bsz} G. B{\"o}m and K. Szlach\"anyj,
Lett. Math. Phys. {\bf 35}, 437 (1996).
\bibitem{powo} P. Podle\'s and S. L. Woronowicz, Commun. Math. Phys.
{\bf 178}, 61 (1996).
\bibitem{mukh}
E. E. Mukhin, Commun. Alg. {\bf 22}, 451 (1994).
\bibitem{dem}
E. E. Demidov, On Wess--Zumino complexes, Doklady AN {\bf 517}, 71
(1993).
\bibitem{WZ}
J. Wess and B. Zumino, Covariant differential calculus
on the quantum hyperplane, CERN Preprint 5697/90
\bibitem{gzh}
A. Giaquinto and J. J. Zhang, J. Algebra {\bf 176}, 861 (1995).
\bibitem{das} C. Daskalsyannis, J. Phys. {\bf A24}, l789, (1991).
\bibitem{jws} P.E.T. J\o rgensen, L.M. Schmitt, and R.F. Werner,
Pac. J. Math. {\bf 165}, 1 (1994).
\bibitem{pie} R. S. Pierce, Associative Algebras, Springer Verlag (1982).
\bibitem{gri} M. Griesl, J. Geom. Phys. {\bf 17}, 90 (1995).
\end{thebibliography}
\end{document}